\documentclass[sigconf,nonacm]{acmart}

\usepackage{amsmath}
\usepackage{braket}          
\usepackage{listings}
\usepackage{xcolor}
\usepackage{graphicx}
\usepackage{booktabs}        
\usepackage{hyperref}
\usepackage{microtype}
\usepackage{cleveref}

\setcopyright{none}
\acmDOI{}
\acmISBN{}
\acmConference[]{}{}{}
\acmYear{2026}
\copyrightyear{2026}


\author{Ana Paula Appel}
\email{aappel@redhat.com}
\affiliation{%
  \institution{Red Hat, Inc.}
  \country{USA}
}

\title{Quantum Encoding Agents: A Natural Language Interface for\\
       Data Embedding Strategy Selection in Quantum Machine Learning}


\begin{document}

\begin{abstract}
Selecting an appropriate data encoding strategy is one of the most consequential
and least-tooled decisions in Quantum Machine Learning (QML) pipelines. The choice
of quantum feature map determines the structure of the Hilbert space into which
classical data is embedded, directly shaping the expressibility of quantum kernels,
the trainability of variational circuits, and the feasibility of execution on
near-term hardware. Despite its central role, encoding selection is rarely addressed
systematically: practitioners typically default to a single strategy---most often
angle encoding---without analyzing how the structural characteristics of their data
interact with the properties of each encoding family. This paper presents
\textbf{Quantum Encoding Agents}, an open-source system that reformulates encoding
selection as a natural language interaction. Given a dataset and an optional
description of the QML task, the system analyzes the data profile, applies a
hardware-aware recommendation policy grounded in the critical gate error threshold
$p^{*} \approx 10^{-3}$, generates a complete copyable Qiskit circuit, produces a
natural language justification in the user's language (Portuguese or English), and
computes the quantum kernel matrix with Kernel-Target Alignment (KTA) scoring---a
label-aware scalar metric that measures how well the quantum kernel groups
same-class samples and separates different-class samples in Hilbert space.
When a numerical matrix is available, the system first estimates the correlation
fractal dimension $D_2$ and selects original columns via FD-ASE~\cite{appel2026fractal,sousa2002fractal},
using $q^{*}=\max(2,\lceil D_2\rceil)$ as an \emph{a priori} qubit budget so that
angle and IQP maps are not recommended at a width where the fidelity kernel has
already collapsed. The system
is implemented as a FastAPI microservice, deployed on Red Hat OpenShift with NVIDIA
OpenShell sandbox isolation, and exposed through OpenClaw agents with distinct
epistemic personalities calibrated to different levels of user expertise. Evaluation
confirms that the system correctly selects among seven encoding families---amplitude,
angle, dense angle, IQP, basis, data re-uploading, and custom feature map---under
realistic NISQ hardware constraints, and that KTA scores discriminate between
encodings on benchmark datasets.
\end{abstract}

\maketitle

\section{Introduction}

The practical adoption of Quantum Machine Learning faces a fundamental bottleneck
that precedes model training: the problem of quantum data encoding. Before any
quantum classifier, kernel method, or variational circuit can operate, classical
data must be mapped into quantum states. This mapping---the encoding or feature
map---is not a neutral preprocessing step. It defines the geometry of the state
space, the structure of correlations captured between features, and the depth and
qubit count of the resulting circuit. Different encodings impose radically different
tradeoffs: amplitude encoding compresses $n$ features into $\lceil \log_2 n \rceil$
qubits but requires a state preparation circuit of exponential depth; angle encoding
produces shallow circuits with one qubit per feature but limited expressibility;
IQP encoding~\cite{havlicek2019} captures cross-feature correlations through
diagonal unitaries and has strong theoretical guarantees for quantum kernels, but at
greater circuit depth.

Despite the critical nature of this choice, the QML literature offers little
guidance on encoding selection as a function of data characteristics. Practitioners
encounter a combinatorial decision problem: seven or more encoding families, each
with different qubit efficiency, circuit depth, hardware sensitivity, and affinity
for different QML algorithms, must be matched against a dataset whose relevant
properties---dimensionality, value type, distribution, sign---interact with the
encoding in non-obvious ways. The seminal survey by Sammartino~\cite{sammartino2026},
reviewing 66 papers from 2017 to 2026, identifies this as the most underaddressed
problem in applied QML, formalizes six encoding families, and derives a critical gate
error rate threshold $p^{*} \approx 10^{-3}$ above which deep encodings become
impractical on NISQ hardware.

This paper addresses the encoding selection problem through a different lens: rather
than proposing a new encoding or a new theoretical framework, I build a practical
agent system that makes existing knowledge actionable. I make the following
contributions:

\begin{enumerate}
  \item \textbf{A hardware-aware encoding recommendation engine} that applies the
    $p^{*}$ threshold from~\cite{sammartino2026}, accounting for gate error rate,
    circuit depth budget, qubit count, and chip connectivity topology.

  \item \textbf{A fractal qubit budget}~\cite{appel2026fractal} that estimates
    $D_2$ (LiBOC) and selects original columns with FD-ASE~\cite{sousa2002fractal}
    \emph{before} the recommendation tree and the labelled KTA search, so
    angle/IQP are not proposed at a width the kernel cannot bear. The crop is
    the default; \texttt{apply\_fractal\_budget=false} still reports $D_2$ and
    the suggested columns while encoding the full table.

  \item \textbf{Seven implemented encoding strategies} in Qiskit, including
    dense-angle encoding~\cite{sammartino2026} and IQP encoding~\cite{havlicek2019},
    which were identified as underrepresented in applied tooling.

  \item \textbf{A natural language explanation layer} that generates structured
    justifications in Portuguese and English, citing concrete metrics (number of
    qubits, circuit depth, Kernel-Target Alignment (KTA), $D_2$, $q^{*}$) derived from actual
    simulation results.

  \item \textbf{A quantum kernel evaluation endpoint} that computes
    $K_{ij} = |\langle\phi(x_i)|\phi(x_j)\rangle|^2$ for a dataset, returns KTA
    scores, ranks encodings when labels are available, and reports the
    operational kernel-alive diagnostic of~\cite{appel2026fractal}
    (near/far fidelities and mean off-diagonal) on the same matrix.
    \texttt{/v1/compare} also sweeps qubit width $q$ under FD-ASE, PCA, and
    CSV-prefix views so $q^{*}$ can be compared with the PCA-95\% allocation.

  \item \textbf{A multi-agent deployment architecture} on Red Hat OpenShift with
    NVIDIA OpenShell security sandboxes, exposing three OpenClaw agents with distinct
    communicative personalities for expert and novice audiences.

  \item \textbf{A Kubeflow Pipeline} for Red Hat OpenShift AI that orchestrates the
    full workflow---data analysis, encoding recommendation, comparison, kernel
    evaluation, and MLflow artifact logging---as a reproducible five-stage DAG.
\end{enumerate}

The system is released as open source.\footnote{
  API: \url{https://github.com/anapaulaappel/quantum-encoding-agents};
  Agents: \url{https://github.com/anapaulaappel/openclaw-quantum-agents}.
  All code, policies, agent configurations, and the Kubeflow pipeline were
  developed and are maintained by the author.}

\section{Related Work}

\subsection{Quantum Data Encoding Strategies}

The theoretical landscape of quantum data encoding has been shaped by a sequence of
foundational papers. Schuld and Killoran~\cite{schuld2019} established the
connection between quantum feature maps and kernel methods, showing that quantum
circuits define kernels on data through the inner product of encoded states in
Hilbert space. This framing---encoding as an implicit kernel---is the basis for all
kernel-based QML methods.

Havl\'i\v{c}ek et al.~\cite{havlicek2019} operationalized this idea with a concrete
circuit construction: the ZZ feature map, which applies Hadamard layers followed by
diagonal Pauli rotations encoding both linear ($x_i^2$) and cross-product
($x_i \cdot x_j$) terms. This circuit, equivalent to what we term IQP
(Instantaneous Quantum Polynomial) encoding, was demonstrated on IBM quantum
hardware and showed experimental advantage over classical SVMs on a constructed
classification task.

Thanasilp et al.~\cite{thanasilp2024} later proved that global quantum kernels
concentrate exponentially in the number of qubits: off-diagonal fidelities vanish
and $K$ becomes the identity. Appel~\cite{appel2026fractal} showed that this
collapse is often \emph{premature} relative to the data: the recorded width $E$ is
an embedding dimension, while the correlation fractal dimension $D_2$ estimates
the intrinsic dimension of the support. Encoding all $E$ columns---or the
PCA-95\% width---in an angle or IQP map can kill the kernel before any label is
seen. On Breast Cancer ($E=30$), a one-layer $ZZ$ fidelity kernel is geometrically
dead by four qubits, while $D_2 \approx 2.5$ yields a three-qubit budget that
stays alive~\cite{appel2026fractal}. FD-ASE~\cite{sousa2002fractal} then chooses
a subset of \emph{original} columns whose partial dimension recovers $D_2$; unlike
PCA, it does not return linear mixtures. The present system operationalizes that
budget: $q^{*}=\max(2,\lceil D_2\rceil)$ caps circuit width \emph{before}
encoding recommendation and KTA ranking.

P\'erez-Salinas et al.~\cite{perezsalinas2020} introduced data re-uploading,
departing from the one-shot encoding paradigm. In their framework, classical data is
inserted into the circuit multiple times across layers, interleaved with trainable
parameters. Data re-uploading has become the de facto encoding for variational
quantum classifiers and quantum neural networks.

LaRose and Coyle~\cite{larose2020} conducted the first systematic comparative study
of encoding strategies across multiple datasets and noise conditions, establishing
that no single encoding dominates and that the optimal choice is dataset-dependent.
Their work motivates the need for an automated selection mechanism---a gap this
system addresses.

The most comprehensive treatment is Sammartino's 2026 survey~\cite{sammartino2026},
which reviews 66 papers from 2017 to 2026, classifies encodings into six families,
and derives practical selection guidelines. Crucially, Sammartino formalizes the
critical gate error threshold: for hardware with gate error rate
$p \geq p^{*} \approx 10^{-3}$, deep encodings (amplitude, IQP, custom feature map)
accumulate noise faster than their expressibility advantage compensates, making
shallow encodings (angle, dense angle, data re-uploading) preferable. Dense-angle
encoding---which packs two features per qubit via
$R_y(x_{2i}) \cdot R_z(x_{2i+1})$---is identified as the most underused encoding
family in applied QML despite its favorable depth-qubit tradeoff. Dense-angle and
standard angle encodings also introduce \emph{relative phases} between qubits (via
$R_z$ or multi-qubit entangling maps); how much classification signal lives in
phase versus amplitude remains under-explored empirically in QML benchmarks---a gap
we note but do not resolve here.

Recent work emphasizes that the classical map applied \emph{before} $U_\phi(x)$
is as consequential as the circuit itself~\cite{larose2020,bond_trading2025,
pqfm_credit2025}. Our implementation applies min-max scaling per feature to
$[0,\pi]$ before rotation-based encodings and reports degeneracy warnings.
In quantum circuits gate order is generally non-commutative; for encoding,
the order in which features are mapped onto qubits can change the implicit
kernel even for a fixed ansatz family~\cite{fioravanti2025_encoding_opt}.
Our implementation (\texttt{encoding\_optimization.py}) runs a three-phase
greedy search---permutation, backward selection, and per-feature
weights---maximizing KTA when labels are present; it is enabled via
\texttt{optimize\_features=true} on \texttt{/v1/compare/csv} and the
\texttt{compare\_csv\_embeddings} tool.
Havl\'i\v{c}ek et al.~\cite{havlicek2019} describe IQP/Hamiltonian encodings
with a specific ZZ Hamiltonian; broader Heisenberg/Ising and Projected Quantum
Feature Maps~\cite{pqfm_credit2025,pqfm_failure2026,quenched_qfm2025} differ in
preprocessing and optional initial states $|s\rangle$---discussed as related
work rather than implemented encodings here.

\subsection{Encoding as an Implicit Kernel}

A result fundamental to the theoretical grounding of this system is Schuld's 2021
theorem~\cite{schuld2021_kernel}: every supervised quantum model that encodes data
into a quantum state $|\phi(x)\rangle$ and measures an observable is mathematically
equivalent to a kernel support vector machine with kernel:
\begin{equation}
  K(x, x') = |\langle\phi(x)|\phi(x')\rangle|^2.
  \label{eq:kernel}
\end{equation}
This equivalence has a decisive practical implication: \emph{``the way that data is
encoded into quantum states is the main ingredient that can potentially set quantum
models apart from classical machine learning models''}~\cite{schuld2021_kernel}.
Furthermore, kernel-based training---which is what QSVM performs directly---is
provably at least as good as variational circuit training for the same encoding. The
choice of encoding is therefore not a preprocessing decision; it is the core
modeling decision, equivalent to choosing a kernel in classical SVM. Product-state
encodings are typically classically simulable; entangled maps such as full-pair IQP
and custom feature maps are heuristically harder to simulate---the comparison report
includes a classical-simulability assessment per encoding. This system makes this
choice explicit and measurable---the Kernel-Target Alignment (KTA) score computed
by \texttt{/v1/kernel} and the multi-encoding KTA ranking in
\texttt{/v1/compare/csv} are direct, label-aware evaluations of the implicit kernel
defined by any encoding on the actual data.

\subsection{Trainability and Barren Plateaus}

The trainability of quantum circuits is a central concern for encoding choices used
in variational algorithms. McClean et al.~\cite{mcclean2018} demonstrated that
randomly initialized quantum circuits exhibit barren plateaus: exponentially
vanishing gradients that make training infeasible beyond moderate qubit counts.
Cerezo et al.~\cite{cerezo2021} surveyed the landscape of variational quantum
algorithms, establishing data re-uploading as the preferred input encoding for QNNs
and VQCs due to its per-layer data injection that maintains gradient flow.

\subsection{Expressibility, Entanglement, and the Encoding Tradeoff}

Sim, Johnson and Aspuru-Guzik~\cite{sim2019} define \emph{expressibility} as the
deviation of a parameterized quantum circuit's output distribution from the Haar
measure. Their key result: shallow circuits have low expressibility and are confined
to structured submanifolds of the state space, which may be insufficient to separate
classes. This is the complementary risk to the barren plateau: too-shallow encodings
underfit in Hilbert space, while too-deep circuits become untrainable due to
vanishing gradients~\cite{mcclean2018}. Sammartino~\cite{sammartino2026} adds the
third axis: above $p^{*} \approx 10^{-3}$, the depth budget is set by hardware
decoherence, not by expressibility requirements.

Larocca et al.~\cite{larocca2023} further establish that over-parameterized quantum
neural networks undergo a phase transition from a barren plateau regime to a
trainable regime as the number of parameters exceeds a critical threshold, directly
linking encoding circuit depth to parameter count and trainability.

\subsection{Quantum Advantage and Scope of NISQ-Era QML}

Huang et al.~\cite{huang2022} proved exponential quantum advantage in a specific
learning task---predicting properties of quantum physical systems from quantum
measurements---demonstrated on 40-qubit superconducting hardware. However, this
advantage is task-specific: it applies when the data itself is quantum in origin,
not when encoding classical tabular data into quantum circuits. Huang, Kueng and
Preskill~\cite{huang2021} establish complementary information-theoretic bounds
showing that classical ML can match quantum ML on average over input distributions,
but quantum advantage is possible worst-case.

For classical datasets on NISQ hardware---the regime this system targets---quantum
advantage over classical methods has not been proven in general. Bowles, Ahmed and
Schuld~\cite{bowles2024} show, in a large-scale study across 12 QML models and 160
datasets, that classical baselines frequently match or outperform quantum classifiers
at current qubit scales, underscoring that encoding evaluation is not optional but
essential. This system is therefore designed as a NISQ-era engineering tool for
informed encoding selection, not as a claim of general quantum supremacy.

\subsection{NISQ vs.\ Fault-Tolerant Quantum Computing}

The \texttt{hardware\_profile} constraints in this system encode NISQ-era realities
that will shift significantly in the fault-tolerant regime. On current NISQ
devices---characterized by gate error rates of $10^{-3}$ to $10^{-2}$, qubit counts
of 10--400, and no error correction---circuit depth is the primary engineering
constraint. In the fault-tolerant era, with logical qubits suppressing error to
arbitrarily low levels, this constraint disappears: amplitude encoding's exponential
qubit compression becomes viable, and the encoding choice is driven purely by
expressibility and task alignment. This design is forward-compatible: passing
\texttt{gate\_error\_rate: 0.0} returns recommendations based on expressibility
alone; passing \texttt{gate\_error\_rate: 5e-3} (IBM Eagle) returns
hardware-adjusted recommendations.

\subsection{AI Agents for Scientific Tooling}

The use of LLM-based agents as interfaces for scientific computing tools is an
emerging paradigm. In quantum computing, recent work spans LLM-assisted architecture
design~\cite{llm_qarch2023}, variational circuit agents~\cite{vqc_agents2026,
llm_vqc_system2026}, QGAN ansatz optimization~\cite{llm_qgan2025}, broader surveys
of AI for quantum computing~\cite{ai_quantum_nature2025,ieee_llm_qc2025}, and
generative circuit synthesis~\cite{diffusion_circuit_synthesis2024}---adjacent to,
but distinct from, encoding selection. \textbf{This paper's scope is encoding
strategy selection and kernel evaluation}, not autonomous circuit generation.

To our knowledge, this is the first open-source system to deploy multi-personality
LLM agents specifically for hardware-aware quantum encoding recommendation with
empirical KTA ranking on user datasets.

The OpenClaw agent framework~\cite{openclaw2026} provides the infrastructure for
agent deployment with persistent memory (via \texttt{MEMORY.md}), user calibration
(via \texttt{BOOTSTRAP.md} and \texttt{USER.md}), and skill encapsulation.

\subsection{Secure Agent Deployment}

NVIDIA OpenShell~\cite{nvidia2026} addresses agent security through out-of-process
policy enforcement: each agent runs in an isolated sandbox governed by a YAML policy
that constrains filesystem access and network endpoints at HTTP method granularity,
enforced by Landlock LSM and seccomp. The policy engine validates rules using the Z3
SMT solver before application. In this deployment, each agent receives a distinct
policy: the expert agent (Circuit) is restricted to the Qiskit API and LLM endpoints
only; the mentor agent (Quanta) additionally has read-only access to arXiv and Qiskit
documentation.

\section{Problem Description}

\subsection{The QML Training Pipeline}

To position the encoding selection problem, I first establish the full QML training
loop in which it occurs. A supervised quantum machine learning pipeline consists of
five stages executed iteratively (Figure~\ref{fig:qml-pipeline}):

\begin{figure}[t]
  \centering
  \includegraphics[width=\linewidth]{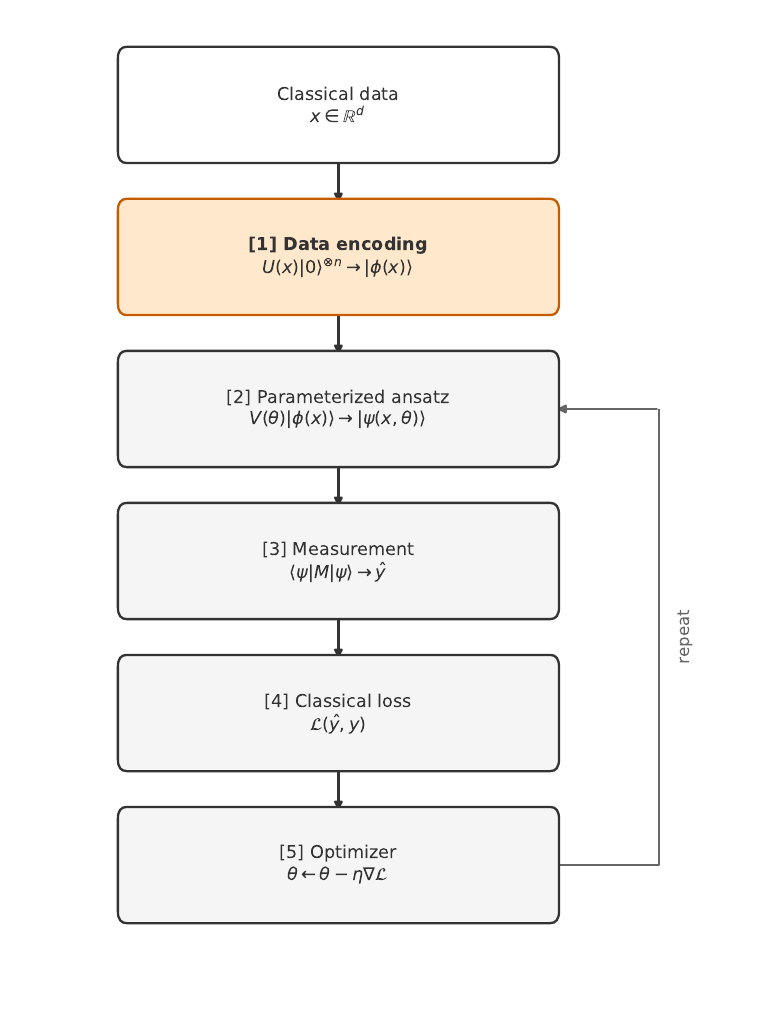}
  \caption{Supervised QML pipeline. Stage~1 (data encoding) is fixed during training
    and is the focus of this paper.}
  \label{fig:qml-pipeline}
\end{figure}

\begin{enumerate}
  \item \textbf{Data Encoding}: $U(x)|0\rangle^{\otimes n} \rightarrow |\phi(x)\rangle$
        \hfill \emph{(this paper)}
  \item \textbf{Parameterized Ansatz}: $V(\theta)|\phi(x)\rangle \rightarrow |\psi(x,\theta)\rangle$
  \item \textbf{Measurement}: $\langle\psi(x,\theta)|M|\psi(x,\theta)\rangle \rightarrow \hat{y}$
  \item \textbf{Classical Loss}: $\mathcal{L}(\hat{y}, y) \rightarrow \mathbb{R}$
  \item \textbf{Classical Optimizer}: $\theta \leftarrow \theta - \eta\nabla\mathcal{L}$
        (then repeat)
\end{enumerate}

The encoding layer (stage 1) is fixed for the duration of training---it is a
hyperparameter of the model, not a trainable parameter. A poorly chosen encoding
defines a kernel that cannot linearly separate the classes in Hilbert space, no
matter how many training iterations the optimizer runs. This system operates at
stage 1, with KTA providing a measurement of whether the chosen encoding creates a
Hilbert space geometry useful for the loss at stage 4.

For quantum kernel methods (QSVM), stages 2--5 are replaced by classical SVM
training on the kernel matrix
$K_{ij} = |\langle\phi(x_i)|\phi(x_j)\rangle|^2$,
computed by the \texttt{/v1/kernel} endpoint. In this case the encoding is the
\emph{entire} model specification, reinforcing the centrality of stage~1.

\subsection{The Encoding Selection Problem}

Let $\mathcal{D} = \{x_i\}_{i=1}^N \subset \mathbb{R}^d$ be a classical dataset
with $N$ samples and $d$ features. A quantum encoding is a parameterized map
$\phi: \mathbb{R}^d \rightarrow \mathcal{H}$ from the feature space to a Hilbert
space $\mathcal{H}$ of dimension $2^n$, implemented as a quantum circuit $U(x)$
acting on $n$ qubits such that $|\phi(x)\rangle = U(x)|0\rangle^{\otimes n}$.

The encoding selection problem is: given $\mathcal{D}$ and an optional QML task
specification $\mathcal{T}$ (classification, kernel method, variational circuit,
etc.)\ and hardware constraints $\mathcal{H}\!w$ (gate error rate, qubit count,
connectivity topology), select an encoding $\phi^{*}$ that maximizes downstream task
performance subject to hardware feasibility.

This problem is ill-posed for two reasons. First, the downstream task performance
depends on the full learning pipeline, not just the encoding; it cannot be evaluated
without training a model. Second, the interaction between data characteristics and
encoding properties is non-linear and high-dimensional.

\subsection{Current Practice and Its Limitations}

In practice, encoding selection is rarely principled. A survey of QML
implementations in the literature reveals that the majority use angle encoding as a
default, regardless of dataset structure. However, this choice ignores:

\begin{itemize}
  \item \textbf{Dimensionality}: for $d > 10$ features, angle encoding requires $d$
    qubits. Dense-angle encoding achieves the same depth with $\lceil d/2 \rceil$
    qubits. More critically, $d$ is the \emph{embedding} width $E$: if the support
    has intrinsic dimension $D_2 \ll E$, an angle or IQP map of width $E$ can
    collapse the fidelity kernel before labels enter~\cite{appel2026fractal,thanasilp2024}.
    PCA-95\% is not a substitute for $D_2$.

  \item \textbf{Task alignment}: for QSVM, encodings without entanglement define
    kernels with limited expressibility. IQP and custom feature map encodings define
    richer kernels.

  \item \textbf{Hardware noise}: above $p^{*} \approx 10^{-3}$, amplitude encoding's
    deep state preparation circuit accumulates noise faster than its qubit compression
    advantage is worth. On IBM Eagle hardware ($p \approx 5 \times 10^{-3}$), the
    effective circuit depth after error accumulation can reduce the signal-to-noise
    ratio below classification threshold. The fractal budget $q^{*}$ and the chip
    limit \texttt{max\_qubits} are independent caps.

  \item \textbf{Data type}: basis encoding, the natural choice for binary or
    categorical data, is systematically ignored when practitioners default to angle
    encoding for all data types.
\end{itemize}

\subsection{The Accessibility Gap}

Beyond the technical selection problem, there is an accessibility gap: the knowledge
required to make principled encoding choices---distributed across a dozen papers
published between 2018 and 2026, requiring familiarity with quantum information
theory, circuit complexity, and NISQ hardware characteristics---is not accessible to
the ML practitioner approaching QML for the first time.

This gap manifests in two distinct user profiles. The \textbf{QML practitioner}
(expert) needs dense, metric-first information: encoding name, circuit depth, qubit
count, KTA score, and a reference to the relevant paper. The \textbf{ML practitioner
approaching QML} (learner) needs progressive disclosure: physical intuition before
mathematics, the Bloch sphere before Dirac notation, a concrete analogy before a
formal definition. Current tools address neither profile. The system I present
bridges this gap through specialized agents calibrated to each profile.

\section{Quantum Encoding Agents: System Design}

\subsection{Architecture Overview}

\begin{figure*}[t]
  \centering
  \includegraphics[width=0.92\textwidth]{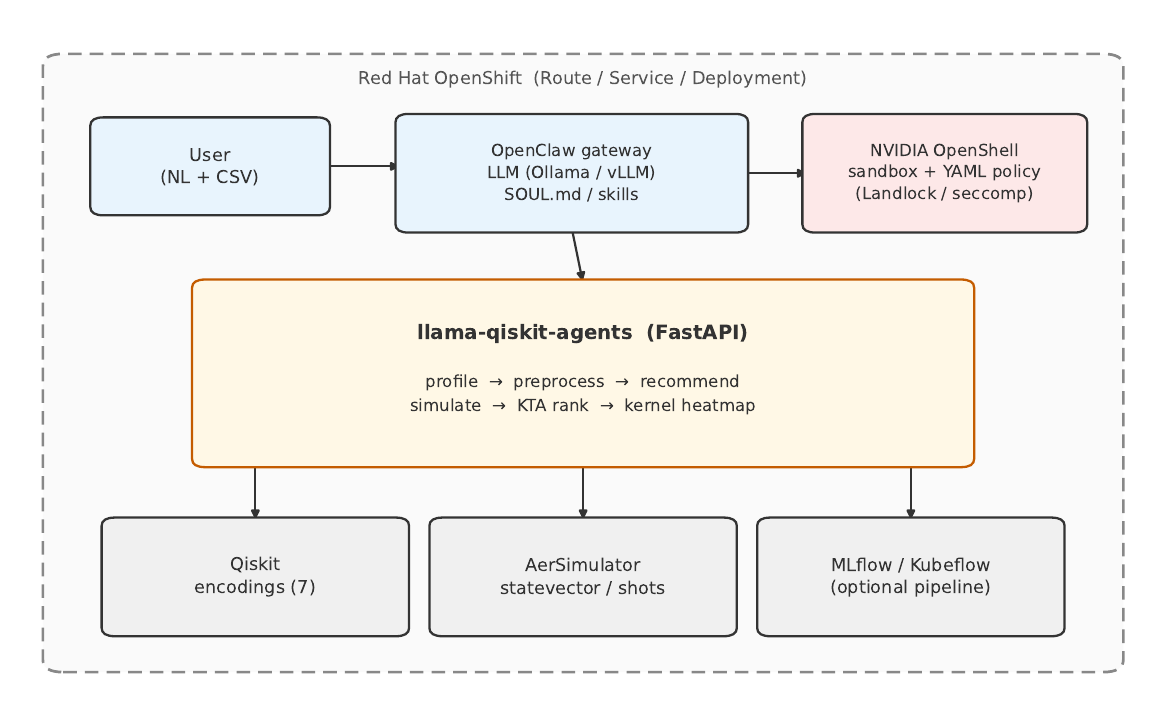}
  \caption{Deployment architecture. Agent gateways call the stateless encoding
    microservice (profile, $D_2$/FD-ASE, recommend, KTA); OpenShell enforces
    per-agent sandbox policies on OpenShift.}
  \label{fig:system-arch}
\end{figure*}

The system consists of three loosely coupled layers (Figure~\ref{fig:system-arch}):
(1)~an OpenClaw agent gateway
with LLM backend (Qwen2.5-Coder-14B via Ollama, or Llama-3.2-3B via vLLM on
OpenShift), (2)~a FastAPI microservice (\texttt{llama\-qiskit\-agents}) exposing
the quantum pipeline, and (3)~the NVIDIA OpenShell security layer enforcing
per-agent sandbox policies. The microservice is stateless and can be called
directly via HTTP independently of the agent framework. Nine quantum tools are
registered under \texttt{/v1/tools} and \texttt{/v1/tools/dispatch}; multi-turn
orchestration is available via \texttt{/v1/agent/chat} and the web UI at
\texttt{/chat} (agent mode with personas or direct CSV comparison without LLM).
OpenClaw and other agent frameworks call the same tool surface over HTTP.

\subsection{Data Profile Analysis}

The first processing step is structured data profiling. Given input in any of three
forms---a numpy array, a CSV file, or a natural language description---the system
extracts a \texttt{DataProfile} dataclass with the structural flags
\texttt{n\_sam\-ples}, \texttt{n\_fea\-tures} (embedding dimension $E$),
\texttt{is\_bi\-na\-ry},
\texttt{is\_cat\-e\-gor\-i\-cal}, \texttt{is\_con\-tin\-u\-ous},
\texttt{has\_neg\-a\-tive}, and \texttt{de\-scrip\-tion}, plus optional fractal
fields ($D_2$, $E$, $q^{*}$, FD-ASE column indices and names,
\texttt{fractal\_selection\_applied}).
For text descriptions, keyword detection in both Portuguese
and English maps phrases to profile flags. Fractal fields stay empty unless a
2-D numeric matrix is present ($N \gtrsim 32$, $E>2$, not binary).

\subsection{The Seven Encoding Strategies}

The system implements seven encoding families as Qiskit \texttt{QuantumCircuit}
objects. Table~\ref{tab:encodings} summarizes their properties.

\begin{table*}[t]
\caption{Quantum encoding strategies implemented in the system.}
\label{tab:encodings}
\setlength{\tabcolsep}{5pt}
\begin{tabular*}{\textwidth}{@{\extracolsep{\fill}}llll p{3.8cm} l@{}}
\toprule
\textbf{Encoding} & \textbf{Qubits} & \textbf{Depth} &
\textbf{Recommended for} & \textbf{Key gates} \\
\midrule
Amplitude       & $\lceil\log_2 n\rceil$ & $O(2^n)$   & Large vectors, qubit-constrained     & \texttt{StatePreparation} \\
Angle           & $d$                    & $1$         & $d\!\leq\!4$, continuous             & $R_y(x_i)$ per qubit \\
Dense Angle~\cite{sammartino2026} & $\lceil d/2\rceil$ & $2$ & $5\!\leq\!d\!\leq\!12$, continuous & $R_y(x_{2i})\,R_z(x_{2i+1})$ \\
IQP~\cite{havlicek2019} & $d$            & ${\sim}3d$  & $8\!\leq\!d\!\leq\!16$, kernels      & $H, R_z(x_i^2), R_{zz}(x_i x_j)$ \\
Basis           & $d$                    & ${\leq}\,d$ & Binary / categorical                 & $X$ per 1-bit \\
Data Re-upl.~\cite{perezsalinas2020} & $d$ & ${\sim}4d$ & VQC, QNN, variational              & $R_y \times L + CX$ chain \\
Custom FM       & $d$                    & ${\sim}3d$  & QSVM, arbitrary kernels              & $H, R_z, R_y, CZ$ pairwise \\
\bottomrule
\end{tabular*}
\end{table*}

\textbf{Dense-angle encoding}~\cite{sammartino2026} achieves half the qubit count of
standard angle encoding at depth 2 by packing two features per qubit:
\begin{equation}
  U_{\mathrm{DA}}(x)\,|0\rangle^{\otimes \lceil d/2 \rceil}
  = \bigotimes_{i=0}^{\lceil d/2 \rceil - 1}
    R_z(x_{2i+1})\,R_y(x_{2i})\,|0\rangle_i.
\end{equation}

\textbf{IQP encoding}~\cite{havlicek2019} is implemented without external
dependencies through manual decomposition of $R_{zz}(\theta) = CX{\cdot}R_z(\theta){\cdot}CX$:
\begin{equation}
  U_{\mathrm{IQP}}(x)
  = H^{\otimes d}
    \cdot \prod_i R_z(x_i^2)
    \cdot \prod_{i<j} R_{zz}(x_i x_j)
    \cdot H^{\otimes d}.
\end{equation}
All feature pairs $(i,j)$ are connected by default; \texttt{pairwise='adjacent'}
restricts $R_{zz}$ to neighboring qubits for linear topologies. Before any circuit,
min-max scaling to $[0,\pi]$ and degeneracy warnings are applied
(\S\ref{sec:preprocess}).

\subsection{Mathematical Preprocessing Pipeline}
\label{sec:preprocess}

Before building $U_\phi(x)$, the system applies encoding-specific classical transforms:
L2 normalization for amplitude encoding; min-max scaling per feature to $[0,\pi]$ for
rotation-based encodings using column bounds from the dataset; and warnings for
near-duplicate angles and NISQ angle resolution. When CSV labels are present,
\texttt{/v1/compare/csv} ranks all encodings by KTA.

\subsection{Fractal Qubit Budget}
\label{sec:fractal-budget}

Angle and IQP encodings spend one qubit per encoded coordinate. If that width
exceeds the intrinsic dimension of the cloud, the fidelity kernel
$K_{ij}=|\langle\phi(x_i)|\phi(x_j)\rangle|^2$ concentrates on the
diagonal~\cite{thanasilp2024,appel2026fractal}. The system therefore computes,
classically and without labels, the correlation fractal dimension $D_2$
(LiBOC box-count~\cite{sousa2002fractal}) and the FD-ASE forward-greedy subset
$E_s$ of \emph{original} columns whose partial dimension recovers $D_2$. The
qubit budget is
\begin{equation}
  q^{*} = \max\bigl(2,\,\lceil D_2\rceil\bigr).
\end{equation}
If $|E_s| > q^{*}$, $E_s$ is truncated to the first $q^{*}$ selected indices:
$q^{*}$ is a hard cap on circuit width. The recommendation tree then uses the
\emph{effective} width $n_{\mathrm{used}}=|E_s|$ (not $E$): angle and IQP are
proposed only when $n_{\mathrm{used}} \le q^{*}$; otherwise the agent prefers
dense-angle or a custom feature map. Generated Qiskit circuits encode only the
selected columns.

This step is unsupervised and runs \emph{before} KTA. The labelled greedy search
of \S\ref{sec:feat-opt} sees the FD-ASE crop, not a fiction of KTA on all $E$
columns. PCA is not used: it would mix axes~\cite{appel2026fractal}.
FD-ASE is skipped for a single row, $E\le 2$, binary data, or a text-only
description. Setting \texttt{apply\_fractal\_budget=false} leaves $D_2$, $q^{*}$,
and \texttt{selected\_columns} on the profile as an advisory, but recommends and
encodes the full width. Chip \texttt{max\_qubits} still applies.

\subsection{Feature Encoding Optimization}
\label{sec:feat-opt}

Following~\cite{fioravanti2025_encoding_opt}, the classical map before
$U_\phi(x)$ includes column order, subset selection, and per-feature weights.
With \texttt{optimize\_features=true} and labels, \texttt{encoding\_optimization.py}
runs a three-phase greedy search maximizing KTA on a fixed encoding \emph{after}
the unsupervised FD-ASE crop (\S\ref{sec:fractal-budget}): exhaustive
or hill-climbing permutation ($d \leq 7$ vs.\ larger $d$), backward column
elimination, and coordinate weight grid $\{0.5,1.0,2.0\}$. The best plan is
applied before min-max scaling and the standard seven-encoding comparison.

\subsection{Hardware-Aware Recommendation Policy}

The recommendation pipeline has three stages: fractal crop (when a matrix is
present), data-driven base selection, and hardware-constrained refinement.

\textbf{Base selection} applies a priority decision tree over the \texttt{DataProfile},
using $n_{\mathrm{used}}$ after FD-ASE (otherwise $E$):
(1)~binary, $n_{\mathrm{used}} \leq 16$ $\to$ \textit{basis};
(2)~continuous, $n_{\mathrm{used}} \leq 4$ and $n_{\mathrm{used}} \le q^{*}$ $\to$ \textit{angle};
(3)~continuous, $4 < n_{\mathrm{used}} \leq 12$, no negatives, dense-angle fits $q^{*}$ $\to$ \textit{dense\_angle};
(4)~single sample, $n_{\mathrm{used}} \geq 4$ $\to$ \textit{amplitude};
(5)~continuous, $8 < n_{\mathrm{used}} \leq 16$ and angle-width $\le q^{*}$ $\to$ \textit{iqp};
(6)~continuous, $n_{\mathrm{used}} > q^{*}$ but dense-angle fits $\to$ \textit{dense\_angle};
(7)~continuous, $n_{\mathrm{used}} > 16$ $\to$ \textit{custom\_feature\_map};
(8)~default $\to$ \textit{angle} only if it fits $q^{*}$.

\textbf{Task refinement} overrides based on the QML task: kernel methods and QSVM
elevate to \textit{custom\_feature\_map}; variational methods prefer
\textit{data\_reuploading}; binary classification with binary inputs uses \textit{basis}.

\textbf{Hardware refinement} applies the $p^{*}$ threshold
from~\cite{sammartino2026}. For any \texttt{HardwareProfile} with
\texttt{gate\_error\_rate} $\geq 10^{-3}$, deep encodings (amplitude,
custom\_feature\_map, iqp) are replaced by \textit{dense\_angle} ($d > 4$) or
\textit{angle} ($d \leq 4$). Heavy-hex or linear connectivity triggers a SWAP
overhead warning for custom\_feature\_map; \texttt{max\_depth\_budget} triggers
a depth feasibility warning.

\subsection{Natural Language Explanation Generation}

The explanation layer generates a four-paragraph structured narrative in the
detected language, each paragraph citing concrete values from the data profile and
simulation result: (1)~data description (including $D_2$, $q^{*}$, and FD-ASE
columns when available), (2)~encoding justification with geometric
intuition, (3)~circuit metrics with NISQ compatibility assessment, and
(4)~comparative analysis of alternatives with concrete depth and qubit differences.
Language detection uses vocabulary overlap between the input text and
Portuguese/English marker sets.

\subsection{Quantum Kernel Evaluation}

For datasets where classification labels are available, the system computes the
quantum kernel matrix $K \in [0,1]^{N \times N}$:
\begin{equation}
  K_{ij} = |\langle\phi(x_i)|\phi(x_j)\rangle|^2.
\end{equation}
Statevectors are computed using the \texttt{StatevectorSimulator} (exact, without
measurement shots) and inner products are computed classically. When class labels
are available, the system additionally reports the \emph{Kernel-Target Alignment}
(\textbf{KTA}): a scalar, label-aware metric, bounded in $[-1, 1]$, that quantifies
how closely the empirical kernel matrix $K$ mirrors an idealized ``target'' kernel
$T = yy^{\top}$ built purely from class membership ($T[i,j]=+1$ if samples $i,j$
share a class, $-1$ otherwise). Formally, it is the (Frobenius) cosine similarity
between the two matrices:
\begin{equation}
  \mathrm{KTA}(K, y)
  = \frac{\langle K,\, yy^{\top} \rangle_F}
         {\|K\|_F \cdot \|yy^{\top}\|_F},
\end{equation}
where $y \in \{-1, +1\}^N$ is the label vector. Intuitively, KTA $\to 1$ indicates
that the quantum feature map places same-class samples close together and
different-class samples far apart in Hilbert space---evidence that the encoding is
likely to yield a separable kernel for QSVM \emph{before} any classifier is trained;
KTA $\to 0$ indicates the kernel carries no information about the classes. The
result is returned as a JSON matrix, a heatmap visualization (viridis colormap,
per-class tick coloring), and a natural language caption.

\begin{figure}[t]
  \centering
  \includegraphics[width=\linewidth]{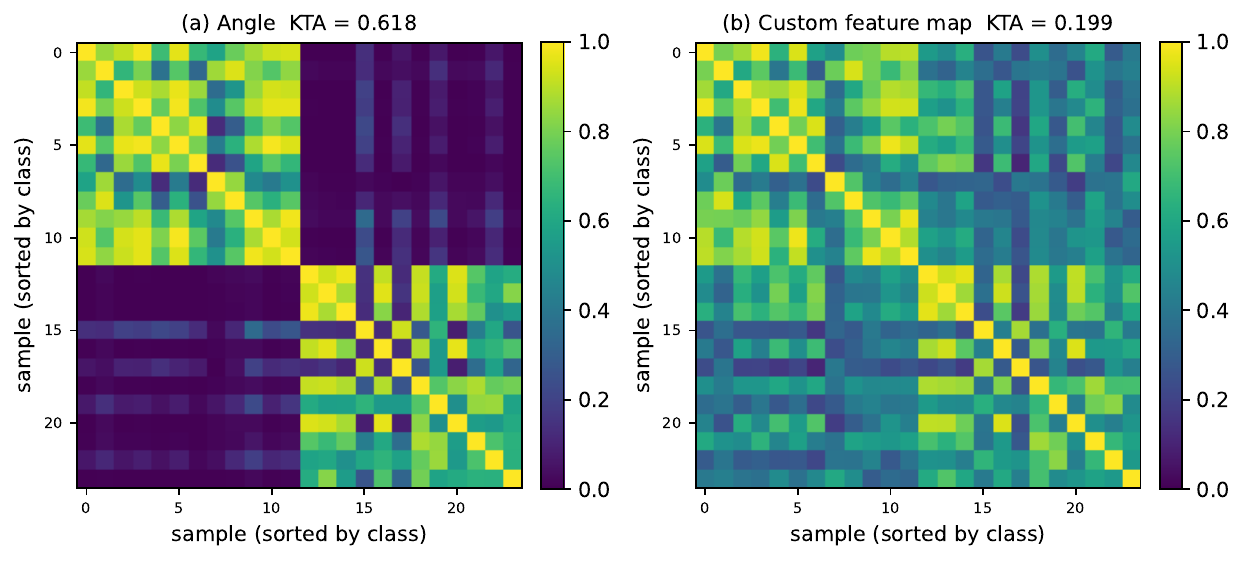}
  \caption{Fidelity kernels on a class-sorted Iris subset
    (12 samples per class; Aer statevector).
    (a)~Angle encoding, subset KTA $= 0.618$.
    (b)~Custom feature map, subset KTA $= 0.199$.
    Full-$N$ KTA values appear in Table~\ref{tab:agent-kta}.
    Simulator only; not a QPU measurement.}
  \Description{Two viridis heatmaps of 24-by-24 kernel matrices. The left panel
    (angle encoding) shows two bright class blocks on the diagonal. The right
    panel (custom feature map) is more mixed, matching its lower KTA.}
  \label{fig:kernel-heatmaps}
\end{figure}

In a preliminary evaluation on a synthetic two-class dataset ($N=4$, $d=2$), angle
encoding achieved KTA $= 0.701$ while custom feature map achieved KTA $= 0.114$,
demonstrating that richer encodings do not necessarily produce better-aligned kernels
and that empirical evaluation via KTA is essential before committing to a training
pipeline.
Figure~\ref{fig:kernel-heatmaps} shows the same contrast on a class-sorted Iris
subset used only for display: block-diagonal structure tracks KTA, while the
subset scores are not identical to the full-$N$ values in
Table~\ref{tab:agent-kta}.
KTA is label-aware. Independently, \texttt{/v1/compare} reports the operational
\emph{kernel-alive} rule of~\cite{appel2026fractal}: the fidelity kernel is alive
when the mean fidelity of Euclidean-nearest pairs is at least $0.25$, the
near/far ratio is at least $2$, and the mean off-diagonal is at least $0.03$.
A high KTA on a dead kernel is a warning that class alignment is being measured
on a nearly diagonal $K$. The same report includes a $q$-sweep of angle encoding
under three views (FD-ASE original columns, PCA, and a CSV prefix) so $q^{*}$
can be compared with the PCA-95\% width that would otherwise over-allocate
qubits.

\subsection{Bloch Sphere and Q-sphere Visualization}

For encodings with $n \leq 6$ qubits, the system optionally captures the statevector
before measurement and renders the Bloch sphere of each qubit's reduced state using
\texttt{plot\_bloch\_\allowbreak multivector}. The resulting PNG (base64-encoded)
accompanies the recommendation when \texttt{include\_bloch:\,true} is set.

\begin{figure*}[t]
  \centering
  \includegraphics[width=0.95\textwidth]{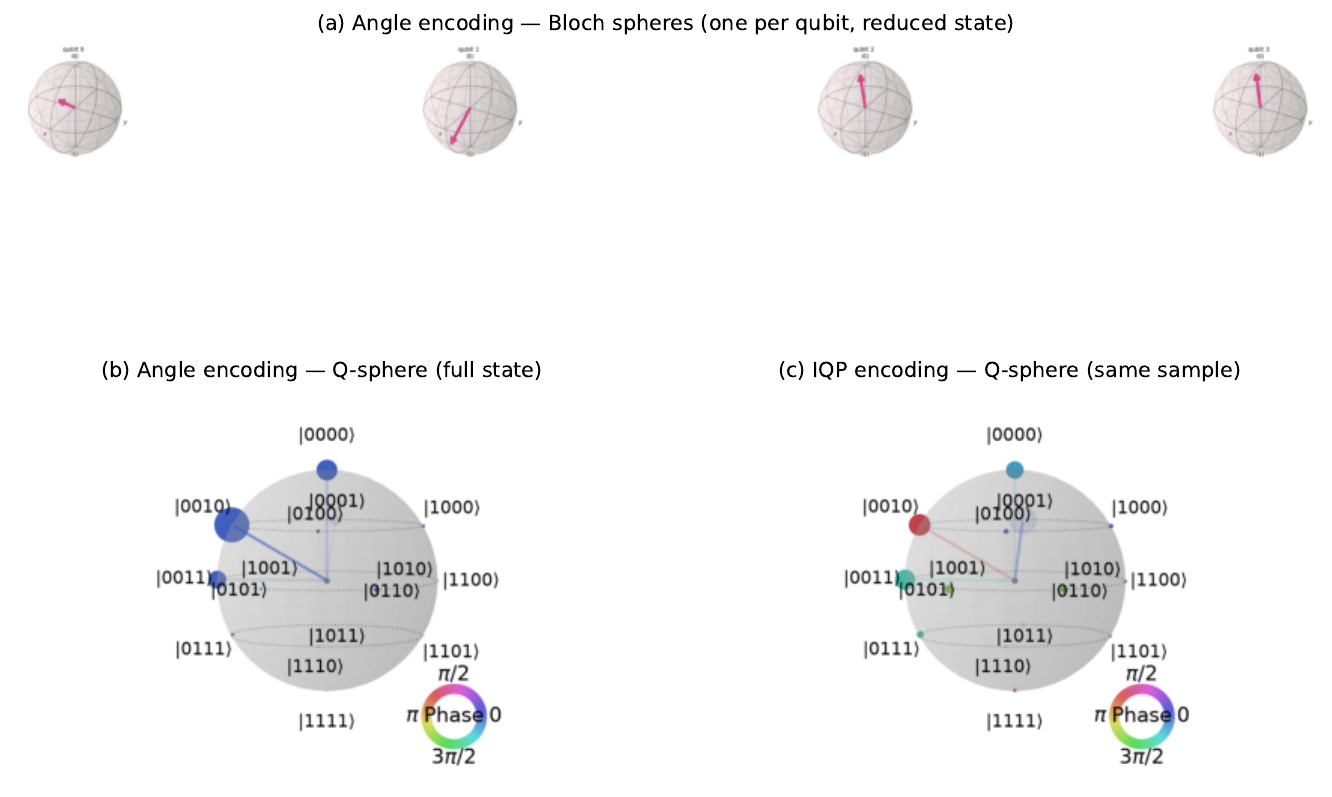}
  \caption{Same Iris sample encoded two ways (Aer statevector).
    (a)~Angle encoding: four independent Bloch vectors (product state).
    (b)~Q-sphere of that product state: amplitudes concentrated, uniform phase.
    (c)~IQP on the same row: phases spread across basis states.
    Simulator only; not a QPU measurement.}
  \Description{Three panels. Top: four Bloch spheres for angle encoding of one
    Iris sample. Bottom left: Q-sphere of the same angle-encoded state with
    mostly single-phase markers. Bottom right: Q-sphere of IQP encoding of the
    same sample with mixed phase colours.}
  \label{fig:bloch-qsphere}
\end{figure*}

The Bloch picture is pedagogically useful but incomplete: each sphere is a
\emph{reduced} one-qubit state, so entanglement is invisible. The Q-sphere
(\texttt{plot\_state\_qsphere}) plots the full $2^n$ amplitudes---latitude by
Hamming weight, marker size by magnitude, colour by phase---so a product-state
encoding and an IQP-style phase map look qualitatively different on the same
sample (Figure~\ref{fig:bloch-qsphere}). The API currently returns Bloch images;
the Q-sphere panels are generated from the same Aer statevector used for kernel
evaluation.

\subsection{Agent Personalities and the Role of SOUL.md}

Three OpenClaw agents expose the recommendation system through distinct communicative
registers. \textbf{QiskitAgent} is the general-purpose agent. \textbf{Circuit}
is calibrated for QML practitioners: its \texttt{SOUL.md} instructs it to
presuppose familiarity with key concepts, respond in three dense blocks (encoding
name, metrics, rationale), never explain superposition, and present trade-offs as
two objective sides without choosing. \textbf{Quanta} is calibrated for learners:
its \texttt{SOUL.md} prohibits the Schr\"odinger's cat analogy (which lies) and
mandates physical intuition before mathematics---the Bloch sphere before Dirac
notation. Both agents perform a calibration ritual on first session
(\texttt{BOOTSTRAP.md}) to populate \texttt{USER.md} with a persistent profile.

\subsection{Skill Architecture}

Two shared skills encapsulate reusable behavior across all agents.
The \texttt{\$qiskit-api} skill documents the recommendation endpoint as a
\texttt{SKILL.md} file, eliminating documentation duplication and ensuring
consistent behavior when the API evolves. The \texttt{\$circuit-review} skill
provides a four-step protocol for reviewing an existing circuit: identify
the encoding, extract metrics, benchmark via \texttt{/v1/compare}, and issue
one of three verdicts (keep / optimize / replace) with a technical rationale.

\subsection{Security Architecture (OpenShell)}

Each agent runs inside a dedicated NVIDIA OpenShell sandbox governed by a YAML
policy validated by the Z3 SMT solver. The policies enforce least-privilege at HTTP
method granularity: Circuit (expert) has access limited to
\texttt{qiskit-api:8080} (read-write) and the LLM endpoint (POST only); Quanta
(mentor) additionally has read-only access to \texttt{arxiv.org} and
\texttt{docs.quantum.ibm.com}; QiskitAgent (general) includes write access to
\texttt{/tmp} for CSV uploads. Binaries are specified as objects with \texttt{path}
keys, enabling per-binary policy enforcement through Landlock LSM.

\subsection{Kubeflow Pipeline for RHOAI}

The full workflow is codified as a Kubeflow Pipelines v2 DAG deployable on Red Hat
OpenShift AI~\cite{rhoai2026}. Five sequential steps orchestrate the pipeline:
(1)~data profiling via \texttt{/v1/analyze},
(2)~encoding recommendation via \texttt{/v1/recommend/explain},
(3)~seven-encoding comparison via \texttt{/v1/compare},
(4)~kernel matrix computation via \texttt{/v1/kernel} (when labels are provided),
and (5)~artifact logging to MLflow~\cite{zaharia2018} (parameters, metrics,
heatmap, Bloch sphere visualization, and generated Qiskit code).
Hardware profile parameters propagate through all steps,
enabling reproducible NISQ-aware experiments.

\section{Discussion}

\subsection{The Role of KTA in Practice}

My preliminary results suggest that KTA is a more reliable pre-training signal than
circuit expressibility alone. Angle encoding (KTA $= 0.701$) outperformed the
custom feature map (KTA $= 0.114$) on the test dataset despite the latter's greater
theoretical richness. This confirms LaRose and Coyle's finding~\cite{larose2020}
that no encoding universally dominates, and motivates the design approach: rather
than recommending based on theoretical properties alone, I compute KTA on the actual
data and present the score alongside the recommendation.

\textbf{Benchmark suite (simulator).}
All KTA values in this section are computed with Qiskit's Aer
\texttt{StatevectorSimulator} via an exact fidelity kernel
$K_{ij}=|\langle\phi(x_i)|\phi(x_j)\rangle|^2$ unless a shot or noise
estimator is named explicitly.
They are \emph{not} QPU counts unless a hardware table says otherwise.
IBM Quantum validation uses the monthly budget in \texttt{benchmarks/HARDWARE.md}
(encoding histograms plus a six-pair kernel-lite; not a full $N\times N$ Gram matrix).
Public sklearn datasets (Iris binary, Breast Cancer top-6 variance, Wine binary,
two moons) and a synthetic order-sensitive set are packaged under
\texttt{benchmarks/}
(\texttt{python scripts/run\_benchmarks.py}).
Shot, noise, and fake-backend transpile experiments are reproduced by
\texttt{paper/figures/generate\_nisq\_experiments.py}.
Table~\ref{tab:feat-opt} reports the greedy order/selection/weight search
(identity CSV order versus the optimized \texttt{FeatureEncodingPlan}).
The \emph{column-order} effect is isolated by the synthetic set, whose CSV
places two noise columns \emph{before} the two class-informative features.

\begin{table}[t]
  \caption{Feature-mapping optimization on the Aer simulator
    (\texttt{scripts/run\_benchmarks.py}, at most 25 kernel samples; not QPU).
    KTA$_{\mathrm{csv}}$ uses the original column order;
    KTA$_{\mathrm{opt}}$ uses the searched plan on the same encoding.}
  \label{tab:feat-opt}
  \centering
  \small
  \begin{tabular}{@{}llrrrr@{}}
    \toprule
    Dataset & Encoding & $N$ & $d$ &
      KTA$_{\mathrm{csv}}$ & KTA$_{\mathrm{opt}}$ \\
    \midrule
    Iris (binary) & angle & 50 & 4 & 0.649 & 0.670 \\
    Synthetic (order) & dense angle & 24 & 4 & 0.553 & 0.674 \\
    Breast Cancer (top-6) & IQP & 40 & 6 & 0.535 & 0.542 \\
    Wine (binary) & IQP & 40 & 13 & 0.204 & 0.550 \\
    \bottomrule
  \end{tabular}
\end{table}

On the synthetic CSV, two noise columns precede the signal
(\texttt{noise\_x}, \texttt{noise\_y}, then \texttt{feat\_a}, \texttt{feat\_b}),
so noise occupies the first qubits.
The optimizer returns
\texttt{feat\_a}~$\to$~\texttt{noise\_y}~$\to$~\texttt{feat\_b}~$\to$~\texttt{noise\_x},
placing an informative $R_y$ at the start of each dense-angle qubit
($\Delta\mathrm{KTA}=+0.121$, about $+22\%$).
On Iris the sklearn order is already near-optimal: the plan keeps
sepal width $\to$ petal length $\to$ petal width and
\emph{drops} sepal length ($\Delta\mathrm{KTA}=+0.021$).
Breast Cancer keeps four of six variance-ranked features
(area error $\to$ worst perimeter $\to$ mean perimeter $\to$ worst texture;
$\Delta\mathrm{KTA}=+0.006$).
Wine shows the largest \emph{selection} gain: IQP on 13 chemical features
collapses to color intensity $\to$ proline
($0.204\to 0.550$, $\Delta\mathrm{KTA}=+0.346$).

\paragraph{Agent choice versus alternatives.}
Table~\ref{tab:agent-kta} compares the data-profile recommendation
(\texttt{/v1/recommend}, the encoding the agent returns unless the user
overrides the task) with three common alternatives, still on the simulator.
On Iris and Breast Cancer the heuristic matches the highest KTA
(angle $0.649$; dense-angle $0.537$).
On the synthetic set the agent picks angle ($0.547$) while dense-angle is
slightly ahead ($0.553$)---a near-tie that KTA ranking can surface.
On Wine the heuristic selects IQP ($0.204$) because $d=13$, but dense-angle
wins KTA ($0.399$); the richer custom feature map ($0.310$) and IQP both
trail the shallow packing.
If the query names a kernel/QSVM task, refinement currently upgrades
\emph{all four} datasets to custom feature map---which is the
\emph{worst} of the four on Iris and synthetic, and only third on Wine.
That gap is why the agent exposes empirical KTA rather than treating the
heuristic as a final answer.

\paragraph{Fractal budget on the full Breast Cancer table.}
On the complete sklearn table ($E=30$),
\texttt{estimate\_fractal\_budget} yields $D_2 \approx 2.49$ and $q^{*}=3$,
matching the \emph{a priori} budget in~\cite{appel2026fractal} (kernel alive near
three qubits; dead at the PCA-95\% width). FD-ASE keeps three original
columns---worst concave points, worst texture, mean fractal
dimension---and \texttt{/v1/recommend} returns \textbf{angle on 3~qubits},
not angle on 30. The default simulator suite (\texttt{breast\_cancer\_fdase})
now scores KTA and kernel-alive on that crop.
Tables~\ref{tab:feat-opt}--\ref{tab:agent-kta} retain the six-column variance
subset used for the original feature-mapping comparison; they are a different
protocol.
With \texttt{apply\_fractal\_budget=false} the same $D_2$ and
column list remain on the profile, but the circuit uses the full width.

\begin{table*}[t]
  \caption{Agent encoding versus alternatives on the Aer statevector
    simulator ($\leq 25$ samples; no QPU). Bold: data-profile recommendation.
    Star: highest KTA among the four encodings shown.}
  \label{tab:agent-kta}
  \centering
  \small
  \begin{tabular}{@{}lccccc@{}}
    \toprule
    Dataset & Angle & Dense-angle & IQP & Custom FM & Agent rec.\ (data) \\
    \midrule
    Iris (binary) & \textbf{0.649}$^\star$ & 0.630 & 0.363 & 0.230 & angle \\
    Synthetic (order) & \textbf{0.547} & 0.553$^\star$ & 0.213 & 0.128 & angle \\
    Breast Cancer (top-6) & 0.526 & \textbf{0.537}$^\star$ & 0.535 & 0.469 & dense-angle \\
    Wine (binary) & 0.335 & 0.399$^\star$ & \textbf{0.204} & 0.310 & IQP \\
    \bottomrule
  \end{tabular}
\end{table*}

\begin{figure}[t]
  \centering
  \includegraphics[width=\linewidth]{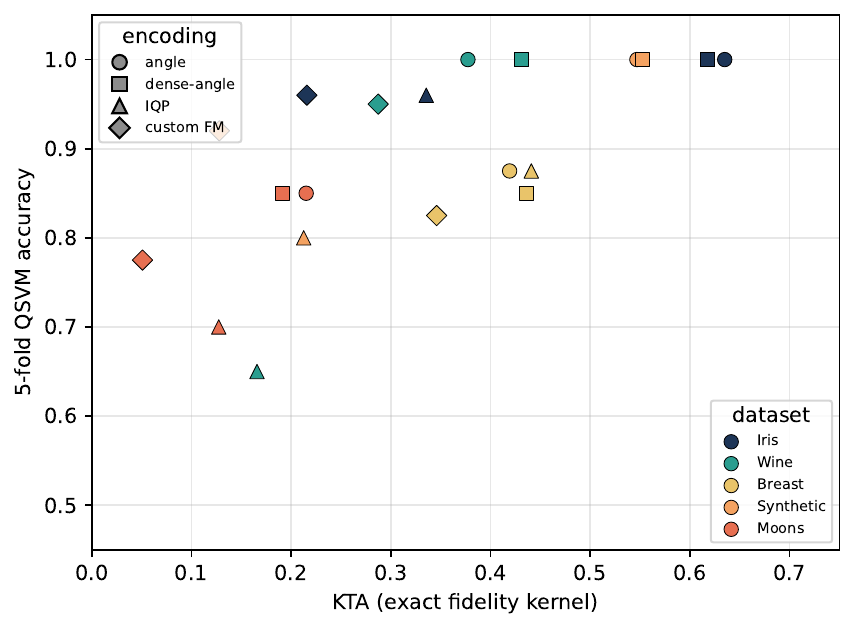}
  \caption{QSVM accuracy versus KTA for the 20 quantum kernels in
    Table~\ref{tab:qsvm} (Aer statevector). Spearman $\rho=0.785$.
    Accuracy saturates on Iris and the synthetic set; moons sit at low KTA.
    Simulator only.}
  \Description{Scatter plot of KTA against 5-fold QSVM accuracy. Points are
    coloured by dataset and shaped by encoding. Higher KTA tends to higher
    accuracy, with a cluster near accuracy 1.0 and a moons cluster at low KTA.}
  \label{fig:kta-acc}
\end{figure}

\paragraph{QSVM versus a classical RBF baseline.}
KTA is a pre-training proxy; Table~\ref{tab:qsvm} reports five-fold stratified
accuracy of an SVM on the precomputed Gram matrix versus sklearn's RBF kernel
(\texttt{gamma='scale'}), still on Aer.
Iris (setosa vs.\ versicolor) and the synthetic set are linearly easy: accuracy
saturates near $1.0$ for RBF and the shallow encodings, so KTA remains the
ranking signal.
Wine ($N=40$) and Breast Cancer ($N=40$) are harder for default RBF
($0.650$ and $0.675$); angle, dense-angle, and (on Breast) IQP all beat RBF.
Two moons ($N=40$, $d=2$, nonlinear) is the opposite check: RBF and the
shallow encodings tie at $0.850$, while IQP and the custom map trail---KTA is
low for every quantum map ($\leq 0.215$).
Across the 20 quantum kernels in Table~\ref{tab:qsvm}, Spearman
$\rho(\mathrm{KTA},\mathrm{acc})=0.785$ (Figure~\ref{fig:kta-acc}).
These are small-$N$ simulator results, not hardware shots.

\begin{table*}[t]
  \caption{5-fold stratified SVM accuracy on precomputed kernels
    (Aer statevector; sklearn \texttt{SVC}; RBF with default
    \texttt{gamma='scale'}). Not QPU.}
  \label{tab:qsvm}
  \centering
  \small
  \begin{tabular}{@{}llccccc@{}}
    \toprule
    Dataset & Metric & RBF & Angle & Dense-angle & IQP & Custom FM \\
    \midrule
    Iris (binary, $N=50$) & Acc. & 1.000 & 1.000 & 1.000 & 0.960 & 0.960 \\
    Iris (binary, $N=50$) & KTA & 0.610 & 0.635 & 0.618 & 0.336 & 0.216 \\
    Wine (binary, $N=40$) & Acc. & 0.650 & 1.000 & 1.000 & 0.650 & 0.950 \\
    Wine (binary, $N=40$) & KTA & 0.163 & 0.377 & 0.431 & 0.166 & 0.287 \\
    Breast (top-6, $N=40$) & Acc. & 0.675 & 0.875 & 0.850 & 0.875 & 0.825 \\
    Breast (top-6, $N=40$) & KTA & 0.158 & 0.419 & 0.436 & 0.441 & 0.346 \\
    Synthetic (order, $N=24$) & Acc. & 1.000 & 1.000 & 1.000 & 0.800 & 0.920 \\
    Synthetic (order, $N=24$) & KTA & 0.120 & 0.547 & 0.552 & 0.212 & 0.128 \\
    Two moons ($N=40$) & Acc. & 0.850 & 0.850 & 0.850 & 0.700 & 0.775 \\
    Two moons ($N=40$) & KTA & 0.196 & 0.215 & 0.191 & 0.127 & 0.051 \\
    \bottomrule
  \end{tabular}
\end{table*}

\begin{figure}[t]
  \centering
  \includegraphics[width=\linewidth]{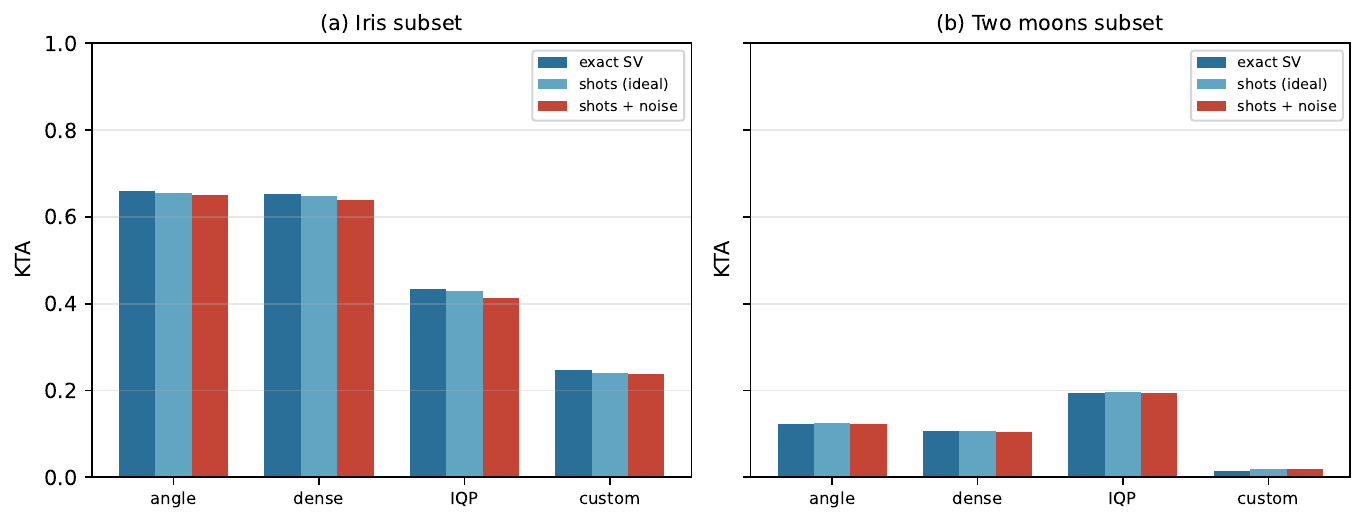}
  \caption{KTA ranking under exact statevectors, ideal shots, and noisy shots
    (same protocol as Table~\ref{tab:shots-noise}).
    Ranking is preserved; moons remain low-KTA for every encoding.}
  \Description{Two grouped bar charts. Left: Iris subset, angle and dense-angle
    near 0.65, IQP near 0.43, custom near 0.24, with exact, shot, and noisy bars
    almost equal. Right: moons subset, all bars below 0.20.}
  \label{fig:shots-noise}
\end{figure}

\paragraph{Shot-based kernels and Aer noise.}
The production endpoint uses an exact statevector fidelity.
A hardware-like estimator is the compute-uncompute overlap
$K_{ij}=\Pr(0^{\otimes n})$ after $U(x_j)^\dagger U(x_i)$, counted from shots.
Table~\ref{tab:shots-noise} and Figure~\ref{fig:shots-noise} compare exact KTA
with this estimator at $256$ shots, then with a depolarizing plus readout
noise model ($p_1=10^{-3}$, $p_2=10^{-2}$, $p_{\mathrm{ro}}=0.02$).
On a stratified $N=16$ slice the encoding ranking is unchanged: mean absolute
error versus the exact Gram matrix is $0.012$--$0.022$ (ideal shots) and
$0.024$--$0.052$ (noisy); noisy diagonals fall to about $0.92$--$0.98$.
KTA itself moves by at most $0.022$. This is still Aer, not a QPU.

\begin{table}[t]
  \caption{Exact fidelity KTA versus compute-uncompute shot kernel
    ($256$ shots) and the same estimator under Aer depolarizing+readout noise.
    Stratified $N=16$ subset. Not QPU.}
  \label{tab:shots-noise}
  \centering
  \small
  \begin{tabular}{@{}llccc@{}}
    \toprule
    Dataset & Encoding & Exact & Shots & Noisy shots \\
    \midrule
    Iris & angle & 0.659 & 0.655 & 0.651 \\
    Iris & dense-angle & 0.653 & 0.649 & 0.640 \\
    Iris & IQP & 0.434 & 0.430 & 0.412 \\
    Iris & custom FM & 0.248 & 0.241 & 0.237 \\
    Moons & angle & 0.123 & 0.124 & 0.122 \\
    Moons & dense-angle & 0.106 & 0.106 & 0.104 \\
    Moons & IQP & 0.194 & 0.196 & 0.194 \\
    Moons & custom FM & 0.015 & 0.018 & 0.019 \\
    \bottomrule
  \end{tabular}
\end{table}

\paragraph{Transpile cost on a fake IBM-like backend.}
Table~\ref{tab:transpile} reports logical depth and two-qubit count versus the
ISA circuit after transpiling one sample to a 57-qubit heavy-hex
\texttt{GenericBackendV2} with IBM-like basis
$\{\texttt{id},\texttt{rz},\texttt{sx},\texttt{x},\texttt{cx}\}$
(optimization level 1; no live device).
Angle and dense-angle stay depth $4$ with zero two-qubit gates.
IQP on Wine ($d=13$) jumps from logical depth $72$ / $156$ two-qubit gates to
ISA depth $190$ / $377$ CNOTs---the same encoding the data-profile heuristic
selects because $d=13$, and the encoding with the \emph{worst} Wine KTA in
Table~\ref{tab:agent-kta}.
Dense-angle on seven qubits remains ISA depth $4$.
The hardware-aware agent should therefore prefer the shallow packing unless
empirical KTA (and a depth budget) say otherwise.

\begin{table*}[t]
  \caption{Logical encoding circuit versus ISA after transpiling to a 57-qubit
    heavy-hex \texttt{GenericBackendV2} (IBM-like basis, optimization level 1).
    One representative sample. Not a live QPU.}
  \label{tab:transpile}
  \centering
  \small
  \begin{tabular}{@{}llrrrrr@{}}
    \toprule
    Dataset & Encoding & $n$ & Depth (log.) & 2q (log.) & Depth (ISA) & 2q (ISA) \\
    \midrule
    Iris ($d=4$) & angle & 4 & 1 & 0 & 4 & 0 \\
    Iris ($d=4$) & dense-angle & 2 & 2 & 0 & 4 & 0 \\
    Iris ($d=4$) & IQP & 4 & 18 & 12 & 36 & 18 \\
    Iris ($d=4$) & custom FM & 4 & 8 & 6 & 32 & 12 \\
    Wine ($d=13$) & angle & 13 & 1 & 0 & 4 & 0 \\
    Wine ($d=13$) & dense-angle & 7 & 2 & 0 & 4 & 0 \\
    Wine ($d=13$) & IQP & 13 & 72 & 156 & 190 & 377 \\
    Wine ($d=13$) & custom FM & 13 & 26 & 78 & 241 & 300 \\
    \bottomrule
  \end{tabular}
\end{table*}

\paragraph{IBM Quantum (ibm\_fez).}
Table~\ref{tab:ibm-hist} reports the four \texttt{week1\_iris} encoding circuits
on IBM Quantum's \texttt{ibm\_fez} (open plan, 512 shots, 9 September 2026):
one sample per class, CSV feature order versus the KTA-optimized plan
(drop sepal length; simulator $\Delta\mathrm{KTA}=+0.023$ on the 12-sample
slice). All four circuits ran as a \textbf{single} Sampler job
(\texttt{dagota8mhr3c73e5fb20}).
Angle encoding transpiles to ISA depth 5 with zero two-qubit gates.
The dominant bitstring matches Aer on every circuit; the hardware peak
probability is 5--11 points lower than Aer, consistent with readout noise
rather than a changed mode.
Table~\ref{tab:ibm-kta} is the kernel-lite that a full QPU Gram matrix cannot
afford: six compute-uncompute overlaps $K_{ij}=\Pr(0^{\otimes n})$ on two
samples per class, submitted as a single Sampler job
(\texttt{daa8m04e74ec73akhtd0}, 30 August 2026).
Intra-class fidelities stay $\approx 0.63$--$0.65$; all four inter-class pairs
remain $\approx 0$. The $4\times 4$ mini-KTA is $0.691$ (exact), $0.693$ (Aer
shots), and $0.688$ on \texttt{ibm\_fez}.
IQP and Wine were not sent to the QPU: Table~\ref{tab:transpile} already shows
hundreds of CNOTs.

\begin{table*}[t]
  \caption{Angle-encoding histograms on IBM \texttt{ibm\_fez} versus Aer
    (512 shots, 9 September 2026). Dominant bitstring agrees; hardware peak
    probability is slightly lower. One Sampler job:
    \texttt{dagota8mhr3c73e5fb20}.}
  \label{tab:ibm-hist}
  \centering
  \small
  \begin{tabular}{@{}llccccc@{}}
    \toprule
    Circuit & $n$ & ISA depth & Mode & $P_{\mathrm{QPU}}$ & $P_{\mathrm{Aer}}$ \\
    \midrule
    Class 0, CSV order & 4 & 5 & \texttt{0010} & 0.670 & 0.783 \\
    Class 0, optimized & 3 & 5 & \texttt{001} & 0.801 & 0.885 \\
    Class 1, CSV order & 4 & 5 & \texttt{1101} & 0.775 & 0.822 \\
    Class 1, optimized & 3 & 5 & \texttt{110} & 0.867 & 0.977 \\
    \bottomrule
  \end{tabular}
\end{table*}

\begin{table}[t]
  \caption{Kernel-lite on the same Iris angle encoding: compute-uncompute
    $K_{ij}$ for two samples per class. Mini-KTA on the filled $4\times 4$
    Gram (diagonal $=1$). One Sampler job on \texttt{ibm\_fez}, 512 shots,
    30 August 2026 (\texttt{daa8m04e74ec73akhtd0}).}
  \label{tab:ibm-kta}
  \centering
  \small
  \begin{tabular}{@{}lccc@{}}
    \toprule
    Pair & Exact & Aer shots & \texttt{ibm\_fez} \\
    \midrule
    Intra class 0 & 0.633 & 0.654 & 0.631 \\
    Intra class 1 & 0.681 & 0.709 & 0.645 \\
    Inter $0_a$--$1_a$ & 0.000 & 0.000 & 0.000 \\
    Inter $0_a$--$1_b$ & 0.004 & 0.002 & 0.004 \\
    Inter $0_b$--$1_a$ & 0.001 & 0.002 & 0.000 \\
    Inter $0_b$--$1_b$ & 0.002 & 0.002 & 0.008 \\
    \midrule
    Mini-KTA ($4\times 4$) & 0.691 & 0.693 & 0.688 \\
    \bottomrule
  \end{tabular}
\end{table}

\subsection{Limits of Text-Based Encoding Selection}

The natural language input path---inferring data profile from a description---has
inherent limitations. Keyword detection can misclassify ambiguous descriptions.
$D_2$ and FD-ASE are not computed on text: without a 2-D array the fractal fields
stay empty and the agent falls back to the previous heuristic. For
production use, numerical data or CSV upload should be preferred; the text path is
primarily useful for initial exploration and for the conversational interface.

\subsection{Agent Personality as an Engineering Artifact}

The \texttt{SOUL.md}-based personality specification is a form of behavioral
engineering that sits between prompt engineering and agent architecture. Unlike a
system prompt, \texttt{SOUL.md} is persistent (injected every session), versioned
in git, and testable: a response from Circuit that begins with ``Great question!''
violates the soul specification and can be detected in evaluation. This framing---
soul as contract---enables more systematic quality assurance of agent behavior than
ad-hoc prompting.

\subsection{OpenShell Policy Design as Security-by-Data-Profile}

The assignment of different network permissions to different agents based on their
communicative role is a novel security design pattern: \textbf{policy as a function
of agent persona}. Quanta requires external internet access (arXiv, Qiskit docs)
because its role is pedagogical and citation-heavy. Circuit does not, because
experts are expected to know the literature. This reflects the insight that the
minimum necessary privilege of an AI agent is determined not just by its technical
function but by its communicative purpose.

\section{Conclusion}

I presented Quantum Encoding Agents, a system that addresses the encoding selection
problem in Quantum Machine Learning through a combination of hardware-aware
recommendation logic, a fractal qubit budget~\cite{appel2026fractal}, natural
language generation, quantum kernel evaluation, and
specialized AI agents calibrated to different user expertise levels. The system is
grounded in the theoretical framework of Sammartino~\cite{sammartino2026},
Havl\'i\v{c}ek et al.~\cite{havlicek2019}, and the $D_2$/FD-ASE result that
angle-encoded kernels collapse when the map is wider than the intrinsic dimension
of the data~\cite{appel2026fractal,thanasilp2024}. It implements seven encoding
families including the underused dense-angle encoding, crops original columns to
$q^{*}$ before KTA, reports kernel-alive beside KTA in \texttt{/v1/compare},
and provides KTA-based encoding quality assessment via a REST API.

The multi-agent architecture demonstrates that AI agents deployed for scientific
tools can and should have distinct communicative personalities calibrated to their
user populations---not as a cosmetic feature, but as a substantive design choice
that determines what information to surface, in what order, and with what degree of
assumed prior knowledge. The security architecture extends this idea: agent policy
is a function of agent role, not just of agent function.

Future work includes: (i)~larger on-device kernels and additional backends
beyond the four encoding circuits and six-pair kernel-lite on
\texttt{ibm\_fez}; (ii)~the Q-Tucker shallow state preparation~\cite{qtuckerarxiv};
(iii)~the Quantum Spectral Model~\cite{qsmarxiv} and spectral
methods~\cite{spectral_schuld2026}; (iv)~Hamiltonian Heisenberg/Ising encodings and
covariant kernels~\cite{glick_covariant2021}; (v)~optional initial states $|s\rangle$
before the feature map as in PQFM literature~\cite{pqfm_credit2025,pqfm_failure2026};
(vi)~extending KTA-guided automatic selection under hardware constraints;
and (vii)~joint optimization of feature plan and encoding family beyond the current
FD-ASE crop and fixed-encoding greedy search~\cite{fioravanti2025_encoding_opt}.

\begin{acks}
The author thanks Jos\'e Victor Soares Scursulim for a careful review of an
earlier draft of this article. His comments substantially improved the treatment
of encoding preprocessing, related work, and mathematical notation.
\end{acks}

\bibliographystyle{ACM-Reference-Format}
\bibliography{references}

\end{document}